\documentclass[12pt]{article}
\usepackage[english]{babel}
\usepackage{amsmath,amsthm,amssymb,amsfonts}
\usepackage{color}
\newtheorem{theorem}{Theorem}

\begin{document}

\title{A thermodynamical description\\ of third grade fluid mixtures}
\author{M.~Gorgone and P.~Rogolino\\
\ \\
{\footnotesize Department of Mathematical and Computer Sciences,}\\
{\footnotesize Physical Sciences and Earth Sciences, University of Messina}\\
{\footnotesize Viale F. Stagno d'Alcontres 31, 98166 Messina, Italy}\\
{\footnotesize mgorgone@unime.it; progolino@unime.it}
}

\date{Published in \textit{J.~Non-Equilib.~Thermodyn.} \textbf{47}, 133--142 (2022).}

\maketitle

\begin{abstract}A complete thermodynamical analysis for a non-reacting binary mixture exhibiting the features of a third grade fluid is analyzed. 
The constitutive functions are allowed to depend on the mass density of the mixture and the concentration of one of the constituents, together with their first and second order gradients, on the specific internal energy of the mixture with its first order gradient, as well as on the symmetric part of the gradient of barycentric velocity. Compatibility with second law of thermodynamics is investigated by applying the extended Liu procedure. An explicit solution of the set of
thermodynamic restrictions is obtained by postulating a suitable form of the constitutive relations for the diffusional mass flux, the  heat flux and the Cauchy stress tensor. Taking a first order expansion in the gradients of the
specific entropy, the expression of the entropy flux is determined. It includes an additional contribution  due to non-local effects.
\end{abstract}


\maketitle

\section{Introduction}
\label{sec:intro}

In the context of rational thermodynamics, a fundamental role is played by models of fluid continua; these models, in many applications, need to be represented by mixtures \cite{Mul1,GurVar,Bow,LiuMul}, \emph{i.e.}, continua  supposed to be microscopically
formed by two or more separate components having different specific physical properties. 

In the general case, for an $N$-component mixture, we have to determine the evolution of  the $N$ partial mass densities $\rho^{(a)}$ $(a=1,\ldots, N)$, the $N$  velocities $\mathbf{v}^{(a)}$, and the $N$ partial temperatures $\theta^{(a)}$ (or, equivalently, partial internal energies 
$\varepsilon^{(a)}$) of the constituents.

Mixtures can be modeled at various degrees of detail \cite{BowGar,PPR2005,GouRug,BotDre,CGOP-2020}. Here, we focus on a  non-reacting model of a binary mixture described by means of the mass density, barycentric velocity and specific internal energy of the mixture, as well as the concentration of one constituent
\cite{GurVar,LiuMul}. More complex models can be considered in the case in which the constituents have different velocities, \cite{BotDre,FraPalRog,FraPalRog1}, or  temperatures \cite{BowGar,GouRug}; recently,  a complete thermodynamical analysis of a binary mixture of viscous Kortweg fluids with two velocities and two temperatures,  has been developed in \cite{GORbis-2021}.

Moreover, in some physical problems, the description of mixtures requires the introduction of internal state variables as 
additional fields \cite{FraPalRog,FraPalRog1,OliPalRog}; this case will not be faced hereafter.

In this paper, we consider a non-reacting mixture exhibiting the features of a third grade fluid, \emph{i.e.}, the constitutive quantities possess second order non-localities \cite{True_Raja}. Solving all the thermodynamical restrictions arising from the entropy principle, we prove that the constitutive equation for Cauchy stress tensor involves the second spatial derivatives of the mass density \cite{DunSer}, \emph{i.e.}, behaves like a Korteweg fluid. 

Korteweg-type fluids have been considered in \cite{Kor}; they are characterized by the following  constitutive equation for the Cauchy stress tensor:
\begin{equation} 
\label{kort}
T_{ij}=\left(-p+\alpha_1\rho_{,k}\rho_{,k}+\alpha_2\rho_{,kk}\right)\delta_{ij}+\alpha_3\rho_{,i}\rho_{,j}+\alpha_4\rho_{,ij},
\end{equation}
where $\rho$ denotes the mass density, $p$ the pressure, $\delta_{ij}$ the Kronecker symbol, and $\alpha_i$ ($i=1,\ldots,4$)  
suitable material functions depending on mass density and temperature
($[\alpha_1]=[\alpha_3]=Kg^{-1}m^7 s^{-2}$, $[\alpha_2]=[\alpha_4]=m^{4}\, s^{-2}$); moreover, the subscripts denote the partial derivatives and the Einstein convention on sums over repeated indices has been used.

These fluids have been investigated in a pioneeristic paper by Dunn and Serrin \cite{DunSer} along  the principles
of rational continuum thermodynamics \cite{TRUE}; moreover, these models have been  extensively studied  by Cimmelli and coworkers  
\cite{CimOliPac3,CimSelTri3,CimSelTri1,CimOlTri,COP-CMT-2015,GOR-2020} through a generalized Liu procedure \cite{Liu,Cim1}, and by 
Heida and M\'alek \cite{HeiMal} following a different methodology. In a recent paper \cite{GP-2021}, an explicit solution of the thermodynamical restrictions imposed by second law of thermodynamics for a viscous Korteweg fluid in the general three-dimensional case has been recovered.

The plan of the paper is as follows. In Section~\ref{sec:balance}, we present the
governing equations for the basic fields we choose to describe the mixture as a whole (mixture mass density, concentration of one constituent, barycentric velocity, specific internal energy), together with the entropy inequality; then, we sketch the extended Liu procedure applied for the exploitation of the entropy principle \cite{CimOlTri,Cim1}, and recover a set of conditions ensuring that second law of thermodynamics is satisfied for 
arbitrary thermodynamical processes. In Section~\ref{sec:3d}, we assume some constitutive relations in order to solve all the thermodynamical restrictions, and provide an explicit, physically admissible, solution. The lengthy computations necessary to derive and solve the algebraic and differential conditions are handled with the help of some symbolic routines written in the Computer Algebra System Reduce \cite{Reduce}. 
As a result, it is proved that the constitutive equations are compatibile with the second law of thermodynamics; in particular, the Cauchy stress tensor for the mixture has been characterized in a very general form typical of Korteweg fluids \cite{GP-2021}. 
Furthermore, by using a semilinear approximation,
phase transitions have been investigated, and a generalized evolution equation for a phase-field, encompassing the classical Cahn-Hilliard equation, has been recovered. Finally, Section~\ref{sec:final} contains our conclusions as well as further 
developments of the present theory.

\section{Basic fields and extended Liu procedure}
\label{sec:balance}
Let us consider a non-reacting binary mixture of third grade fluids, and choose the fundamental fields to describe the mixture as a whole: the mass density $\rho(t,x_j)$ of the mixture, the 
concentration $c(t,x_j)$ of one constituent \cite{GurVar,LiuMul}, the barycentric velocity $\mathbf{v}(t,x_j)\equiv(v_1,v_2,v_3)$, and the specific internal energy $\varepsilon(t,x_j)$ of the mixture. The  mass density of the whole mixture and the concentration of one constituent are given by 
\begin{equation}
\begin{aligned}
&\rho=\rho^{(1)}+\rho^{(2)}, \qquad c=\frac{\rho^{(1)}}{\rho},
\end{aligned}
\end{equation}
where the superscript ${}^{(a)}$ ($a=1,2$) labels the two components.

Therefore, in the absence of external forces and  heat sources, we are led to consider the following governing equations for our mixture, in components: 
\begin{equation}
\label{modelequations}
\begin{aligned}
&\rho_{,t}+\rho_{,j}v_j+\rho v_{j,j}=0,\\
&\rho\left(c_{,t}+v_{j}c_{,j}\right)+J^{(m)}_{j,j}=0,\\
&\rho\left(v_{i,t}+v_j v_{i,j}\right)-T_{ij,j}=0,\\
&\rho\left(\varepsilon_{,t}+v_{j}\varepsilon_{,j}\right)-T_{ij}v_{i,j}+q_{j,j}=0,
\end{aligned}
\end{equation}
where $J^{(m)}_{j}$, $T_{ij}$ and $q_j$ denote the components of the diffusional mass flux, the symmetric Cauchy stress tensor and the heat flux for the  mixture, respectively. 

Then, in our general framework, in order to ensure the compatibility of this model with the second law of thermodynamics, we have to impose that the local entropy production for the mixture,
\begin{equation}
\label{entropyinequality}
\rho \left(s_{,t}+v_j s_{,j}\right)+J^{(s)}_{j,j}\ge 0,
\end{equation}
be non-negative along any admissible thermodynamic process,
where $s$ is the specific entropy, and $J_j^{(s)}$ are the components of the entropy flux.

The field equations (\ref{modelequations}) and the entropy inequality (\ref{entropyinequality}), once the variables entering the state space have been assigned, must be 
closed by the constitutive equations for the diffusional mass flux,  Cauchy stress tensor, heat flux, specific entropy and 
entropy flux. Constitutive theories require that all the constitutive functions must depend on state variables; consequently, let us assume the state space, containing second order non-localities, to be spanned by
\begin{equation}
\label{statespace}
\mathcal{Z}=\left\{\rho,c,\varepsilon,\rho_{,j},c_{,j},L_{ij},\varepsilon_{,j},\rho_{,ij},c_{,ij}\right\},
\end{equation}
where $\displaystyle L_{ij}=\frac{v_{i,j}+v_{j,i}}{2}$ is the symmetric part of the barycentric velocity gradient.

Since we need to find a set of conditions which are at least sufficient for the fulfillment of the entropy inequality (\ref{entropyinequality}) for arbitrary 
thermodynamical processes \cite{CimJouRugVan,JCL}, we apply the extended Liu procedure 
\cite{CimOlTri,Cim1}, incorporating new restrictions consistent with higher order non-local constitutive theories.
According to this procedure, in the entropy inequality, we have to use as constraints the field equations, and their gradient extensions too, up to the order of the gradients entering the state space, by means of suitable Lagrange multipliers. This methodology is motivated by the fact that thermodynamical processes are solutions of the field equations, and, if these solutions are smooth enough, 
are trivially solutions of their differential consequences \cite{RogCim}. 
Since the entropy inequality (\ref{entropyinequality}) has to be satisfied in arbitrary smooth processes,  then, from a mathematical viewpoint, we have to use also the differential consequences of the equations governing those 
processes as constraints for such an inequality. On the contrary, if we limit ourselves to consider as constraints 
only the field equations, we are led straightforwardly to a specific entropy, Lagrange multipliers and Cauchy stress tensor which do not 
depend upon the gradients of the variables entering the state space \cite{CGOP-2020,GOR-2020,GP-2021}, and, hence, losing the typical features of Korteweg fluids.

In order to exploit second law of thermodynamics, we take into account the constraints 
imposed on entropy inequality by the field equations and their gradient extensions; this task is accomplished by introducing the Lagrange multipliers $\lambda^{(1)}$,  $\lambda^{(2)}$, $\lambda^{(3)}_i$, $\lambda^{(4)}$, $\Lambda^{(1)}_i$, $\Lambda^{(2)}_i$,
$\Lambda^{(3)}_{ij}$, $\Lambda^{(4)}_{i}$, $\Theta^{(1)}_{ij}$ and $\Theta^{(2)}_{ij}$ ($i,j=1,2,3$), depending on the state space variables.
Thus, the entropy inequality writes
\begin{equation}
\label{entropyconstrained}
\begin{aligned}
&\rho \left(s_{,t}+v_j s_{,j}\right)+J^{(s)}_{j,j}-\left(\lambda^{(1)}+\Lambda^{(1)}_i\frac{\partial}{\partial x_i}+\Theta^{(1)}_{ik}\frac{\partial^2}{\partial x_i \partial x_k}\right)\left(\rho_{,t}+\rho_{,j}v_j+\rho v_{j,j}\right)\\
&\quad-\left(\lambda^{(2)}+\Lambda^{(2)}_i\frac{\partial}{\partial x_i}+\Theta^{(2)}_{ik}\frac{\partial^2}{\partial x_i \partial x_k}\right)\left(\rho\left(c_{,t}+v_{j}c_{,j}\right)+J^{(m)}_{j,j}\right)\\
&\quad-\left(\lambda^{(3)}_i+\Lambda^{(3)}_{ik}\frac{\partial}{\partial x_k}\right)\left(\rho\left(v_{i,t}+v_j v_{i,j}\right)-T_{ij,j}\right)\\
&\quad-\left(\lambda^{(4)}+\Lambda^{(4)}_{k}\frac{\partial}{\partial x_k}\right)\left(\rho\left(\varepsilon_{,t}+v_{j}\varepsilon_{,j}\right)-T_{ij}v_{i,j}+q_{j,j}\right)\geq 0,
\end{aligned}
\end{equation}
that needs to be expanded by means of the chain 
rule; we perform the task by using some routines written in the Computer Algebra System 
Reduce \cite{Reduce}, and, due to the long computations, we omit to report the full form  here. The main advantages of using symbolic computation are that we are able to extract 
the coefficients of a multivariate polynomial in some derivatives of the field variables, 
and then solve, with the help of the Crack package \cite{Wolf}, the set of differential 
and algebraic conditions for the unknown constitutive functions. 

When the inequality  (\ref{entropyconstrained}) is expanded, we can distinguish the \emph{highest derivatives} and the \emph{higher derivatives}  \cite{CimSelTri1}. 
The formers are both the time derivatives of the field variables and of the
elements of the state space, which cannot be expressed through the governing equations as functions of the 
thermodynamical variables, and the spatial derivatives whose order is the highest one. On the contrary,  the latters are the spatial derivatives whose order is not maximal but higher than that of the gradients entering the 
state space. Due to specific choice of  
the state space (\ref{statespace}), the components of the highest derivatives consist of
\begin{equation}
\begin{aligned}
\mathbf{X}=&\left\{\rho_{,t},c_{,t},v_{i,t},\varepsilon_{,t},\rho_{,it},c_{,it},v_{i,kt},\varepsilon_{,kt},\rho_{,ikt},c_{,ikt},v_{i,jk\ell m},
\varepsilon_{,ijk\ell},\rho_{,ijk\ell m},c_{,ijk\ell m}\right\},
\end{aligned}
\end{equation}
whereas the components of the higher ones are
\begin{equation}
\mathbf{Y}=\left\{\rho_{,jk\ell},c_{,jk\ell},v_{i,jk},\varepsilon_{,jk},\rho_{,ijk\ell},c_{,ijk\ell},v_{i,jk\ell},\varepsilon_{,ijk}\right\}.
\end{equation}
By lengthy though straightforward computation, it is easily
ascertained that the entropy inequality (\ref{entropyconstrained}) can be written in the compact form
\begin{equation}
\label{entropycompatta}
A_p X_p+B^{(3)}_{qrs}Y_qY_rY_s+B^{(2)}_{qr}Y_qY_r+B^{(1)}_qY_q+C\ge 0,
\end{equation}
where the functions $A_p$, $B^{(1)}_{q}$, $B^{(2)}_{qr}$, $B^{(3)}_{qrs}$ and $C$ depend at most on the field and state variables; this means that the left hand side of the inequality (\ref{entropycompatta}) is a polynomial which is linear in the highest derivatives and cubic in the higher ones.
Because of the constraints we imposed,  the highest and higher derivatives may assume arbitrary values \cite{CimOlTri}.

Since in principle nothing prevents the possibility of 
a thermodynamical process where $C=0$, and in order the inequality \eqref{entropycompatta} to be satisfied for arbitrary $X_{p}$ and $Y_q$, 
the conditions 
\begin{equation}
A_p=0, \quad B^{(3)}_{qrs}=0, \quad B^{(1)}_{q}=0,\quad C\ge 0,
\end{equation} 
as well the requirement that the quantities $B^{(2)}_{qr}$ are the entries of a positive semidefinite  symmetric matrix, are sufficient for the fulfillment of the entropy inequality.
 
From $A_p=0$, we obtain the expressions for the components of Lagrange multipliers,
\begin{equation}
\label{lagrange}
\begin{aligned}
&\lambda^{(1)}=\rho\frac{\partial s}{\partial\rho},\qquad\lambda^{(2)}=\frac{\partial s}{\partial c}-\frac{\rho_{,k}}{\rho}\left(\frac{\partial s}{\partial c_{,k}}-2\frac{\rho_{,i}}{\rho}\frac{\partial s}{\partial c_{,ik}}\right)-\frac{\rho_{,ik}}{\rho}\frac{\partial s}{\partial c_{,ik}},\qquad\lambda^{(3)}_i=-\frac{\rho_{,k}}{\rho}\frac{\partial s}{\partial v_{i,k}},\\
&\lambda^{(4)}=\frac{\partial s}{\partial \varepsilon}-\frac{\rho_{,k}}{\rho}\frac{\partial s}{\partial \varepsilon_{,k}}, \qquad\Lambda^{(1)}_k=\rho\frac{\partial s}{\partial\rho_{,k}},\qquad\Lambda^{(2)}_k=\frac{\partial s}{\partial c_{,k}}-2\frac{\rho_{,i}}{\rho}\frac{\partial s}{\partial c_{,ik}}, \\
&\Lambda^{(3)}_{ik}=\frac{\partial s}{\partial v_{i,k}},\qquad
\Lambda^{(4)}_{k}=\frac{\partial s}{\partial \varepsilon_{,k}},\qquad\Theta^{(1)}_{ik}=\rho\frac{\partial s}{\partial\rho_{,ik}},\qquad \Theta^{(2)}_{ik}=\frac{\partial s}{\partial c_{,ik}},
\end{aligned}
\end{equation}
as well as the following restrictions involving the specific entropy and the diffusional mass flux:
\begin{equation}
\label{restricthighest}
\begin{aligned}
&\left\langle\frac{\partial s}{\partial c_{,k\ell}}\frac{\partial J^{(m)}_j}{\partial v_{n,i}}\right\rangle_{(ijk\ell)}=0,&& \left\langle\frac{\partial s}{\partial c_{,k\ell}}\frac{\partial J^{(m)}_j}{\partial \varepsilon_{,i}}\right\rangle_{(ijk\ell)}=0,\\
&\left\langle\frac{\partial s}{\partial c_{,k\ell}}\frac{\partial J^{(m)}_j}{\partial \rho_{,in}}\right\rangle_{(ijk\ell n)}=0,&& \left\langle\frac{\partial s}{\partial c_{,k\ell}}\frac{\partial J^{(m)}_j}{\partial c_{,in}}\right\rangle_{(ijk\ell n)}=0,
\end{aligned}
\end{equation}
where the symbol $\langle\mathcal{F}\rangle_{(i_1\ldots i_r)}$ denotes the symmetric part of the tensor function $\mathcal{F}$ with respect to the indices 
$i_1\ldots i_r$. 

The further thermodynamical restrictions, even if their computation is straightforward, are omitted since their expression is rather long.
Furthermore, by analyzing the conditions which vanish the linear terms in the higher and highest derivatives, \emph{i.e.}, $B^{(1)}_{q}=0$ and $B^{(3)}_{qrs}=0$, it follows that the entropy flux is no longer given by the ratio between the heat flux and the absolute temperature of the mixture, and extra-terms can be obtained. 

In the next section, all the thermodynamical constraints recovered by applying the extended Liu procedure are solved, and explicit constitutive equations for the diffusional mass flux, the Cauchy stress tensor, the heat flux, the specific entropy and the entropy flux are determined. The achieved results provide a complete and physically admissible solution.

\section{A solution of thermodynamical restrictions}
\label{sec:3d}
All the thermodynamical restrictions, derived performing the extended Liu procedure, result too much general for being useful in concrete applications; therefore, a further simplification is necessary in order to  exploit the entropy inequality. Under the hypothesis that the non-reacting binary  mixture of third grade fluids behaves as a viscous Korteweg fluid, let us choose the following ansatz for the Cauchy stress tensor
\begin{equation}
\label{consttensor}
\begin{aligned}
T_{ij}&=\left(\tau_0+\tau_1 \rho_{,k}\rho_{,k}+\tau_2 \rho_{,k}c_{,k}+\tau_3 c_{,k}c_{,k}+\tau_4 \rho_{,kk}+\tau_5 c_{,kk}+\tau_6v_{k,k}\right)\delta_{ij} 
\\
&+\tau_7\rho_{,i}\rho_{,j}+\tau_8\frac{\rho_{,i}c_{,j}+\rho_{,j}c_{,i}}{2}
+\tau_9c_{,i}c_{,j} +\tau_{10}\rho_{,ij}+\tau_{11}c_{,ij}+\tau_{12}L_{ij};
\end{aligned}
\end{equation}
moreover, let us consider the following  constitutive relations  for the diffusional mass and heat fluxes:
\begin{equation}
\label{constJQ}
\begin{aligned}
&J^{(m)}_j=J_{\rho}\rho_{,j}+J_{c}c_{,j},\qquad q_{j}=q\varepsilon_{,j},\\
\end{aligned}
\end{equation}
where $\tau_k$ ($k=0,\ldots,12$), $J_{\rho}$, $J_{c}$, and $q$ are suitable scalar material functions depending at most on $(\rho,c,\varepsilon)$.

In addition, let us expand the specific entropy $s$
around the homogeneous equilibrium state (where all gradients vanish), at the first order on the gradients of mass density of the whole mixture and concentration of one constituent, \emph{i.e.},
\begin{equation}
\label{specificentropy}
s=\hat{s}_0+\hat{s}_1\rho_{,k}\rho_{,k}+\hat{s}_2\rho_{,k}c_{,k}+\hat{s}_3c_{,k}c_{,k},
\end{equation}
where $\hat{s}_i\equiv \hat{s}_i(\rho,c,\varepsilon)$ $(i=0,\ldots,3)$ are some 
scalar functions of the indicated arguments. We are aware that expression (\ref{specificentropy}) is not the most general 
representation of
the entropy density as an isotropic scalar function, but it is enough for our purposes.
\begin{theorem}
The non-local constitutive equations (\ref{consttensor})-(\ref{specificentropy}), associated to 
the system of balance laws (\ref{modelequations}), are compatible with the second law of thermodynamics, i.e., they do not violate the entropy inequality (\ref{entropyinequality}), provided that some physically meaningful conditions on the coefficients therein involved are satisfied.
\end{theorem}
\begin{proof}
On the basis of the assumptions on constitutive relations \eqref{consttensor}-(\ref{specificentropy}), we immediately note that restrictions (\ref{restricthighest}) are identically satisfied; the remaining conditions ($B^{(1)}_{q}=0$, $B^{(3)}_{qrs}=0$)
provide a large set of partial differential equations that we are able to solve
using some routines written in the Computer Algebra System
Reduce \cite{Reduce}. Firstly, the coefficients entering the specific entropy and the diffusional mass flux are obtained:
\begin{equation}
\label{solentropy}
\begin{aligned}
&\hat{s}_0=s_{01}+J_{c}s_{02}+s_{03}, \qquad \hat{s}_1=\frac{J_{\rho}^2-\kappa_2^2}{J_{c}^2}\hat{s}_3+\frac{\phi^\prime}{\rho},\qquad\hat{s}_2=2\frac{J_{\rho}}{J_{c}}\hat{s}_3,\qquad
\hat{s}_3=\frac{s_3}{\rho},\\
&J_{\rho}=\kappa_1 c+\kappa_2,\qquad J_{c}=-\kappa_1\rho+\kappa_3,
\end{aligned}
\end{equation}
with the prime ${}^\prime$ denoting differentiation with respect to the argument,
along with the functions $s_{01}\equiv s_{01}(\varepsilon)$, $s_{02}\equiv s_{02}\left(\frac{J_{\rho}}{J_{c}}\right)$, $s_{03}\equiv s_{03}(\rho)$, $\phi\equiv \phi(\rho)$,
$\kappa_i$  ($i=1,2,3$) and $s_3$ constant. In order to guarantee the principle of maximum entropy at 
equilibrium the quadratic part in the gradients of (\ref{specificentropy}) must be negative semidefinite, whence
\begin{equation}
\label{maxentropy}
\begin{aligned}
&\frac{\kappa_2^2 s_3}{(\kappa_3-\kappa_1\rho)^2}-\phi^\prime\geq 0,\qquad s_3\leq 0.
\end{aligned}
\end{equation}
As far as the absolute temperature $\theta$ of the mixture is concerned, we have
\begin{equation}\label{theta}
\frac{1}{\theta}=\frac{\partial \hat{s}_{0}}{\partial \varepsilon}=\frac{d s_{01}}{d\varepsilon};
\end{equation} 
with the hypothesis that $\displaystyle\frac{d^2s_{01}}{d\varepsilon^2}\neq 0$, by using the implicit function theorem, it follows that the specific internal energy $\varepsilon$ depends only on the absolute temperature $\theta$ of the whole mixture, thus the heat flux takes the form
\begin{equation}
q_j=q\frac{d\varepsilon}{d\theta}\theta_{,j},
\end{equation}
\emph{i.e.}, we recognize the classical Fourier law of heat conduction, wherein the coefficient $\displaystyle q\frac{d\varepsilon}{d\theta}$ is the opposite of the thermal conductivity.

Moreover, we also obtain the following expressions for the material functions entering the Cauchy stress tensor:
\begin{align}\label{T-coeff}
\tau_0&=\rho^2\frac{\partial \hat{s}_0}{\partial \rho}\left(\frac{\partial \hat{s}_0}{\partial \varepsilon}\right)^{-1}=\rho^2\theta\left(-\kappa_1\left(s_{02}-\frac{J_{\rho}}{J_{c}}s_{02}^\prime\right)+s_{03}^\prime\right),\allowdisplaybreaks\nonumber\\ 
\tau_1&=-\frac{\partial \left(\rho^2\hat{s}_1\right)}{\partial \rho}\left(\frac{\partial \hat{s}_0}{\partial \varepsilon}\right)^{-1}=-\theta\left(\frac{(J_{\rho}^2-\kappa_2^2)(J_{c}+2\kappa_1\rho)}{J_{c}^3}s_3+\frac{d(\rho\phi^\prime)}{d\rho}\right), \allowdisplaybreaks\nonumber\\
\tau_2&=-\left(\frac{\partial \left(\rho^2\hat{s}_2\right)}{\partial \rho}+\rho^2\frac{\partial \hat{s}_1}{\partial c}\right)\left(\frac{\partial \hat{s}_0}{\partial \varepsilon}\right)^{-1}=-2\theta\frac{J_{\rho}}{J_{c}^2}(J_{c}+2\kappa_1\rho)s_3, \allowdisplaybreaks\nonumber\\
\tau_3&=-\rho\left(\rho\frac{\partial \hat{s}_2}{\partial c}+\hat{s}_3\right)\left(\frac{\partial \hat{s}_0}{\partial \varepsilon}\right)^{-1}=-\theta\frac{J_{c}+2\kappa_1\rho}{J_{c}}s_3, \allowdisplaybreaks\\
\tau_4&=-2\rho^2 \hat{s}_1\left(\frac{\partial \hat{s}_0}{\partial \varepsilon}\right)^{-1}=-2\rho\theta\left(\frac{J_{\rho}^2-\kappa_2^2}{J_{c}^2}s_3+\phi^\prime\right),\allowdisplaybreaks \nonumber\\
\tau_5&=-\rho^2 \hat{s}_2\left(\frac{\partial \hat{s}_0}{\partial \varepsilon}\right)^{-1}=-2\rho\theta\frac{J_{\rho}}{J_{c}}s_3,\qquad
\tau_7=2\rho \hat{s}_1\left(\frac{\partial \hat{s}_0}{\partial \varepsilon}\right)^{-1}=2\theta\left(\frac{J_{\rho}^2-\kappa_2^2}{J_{c}^2}s_3+\phi^\prime\right),\allowdisplaybreaks\nonumber \\
\tau_8&=2\rho \hat{s}_2\left(\frac{\partial \hat{s}_0}{\partial \varepsilon}\right)^{-1}=4\theta\frac{J_{\rho}}{J_{c}}s_3,\qquad
\tau_9 =2\rho \hat{s}_3\left(\frac{\partial \hat{s}_0}{\partial \varepsilon}\right)^{-1}=2\theta s_3,\qquad
\tau_{10}=\tau_{11}=0,\nonumber
\end{align}
and it is recovered the following form for the entropy flux 
\begin{equation}\label{entropyflux}
\begin{aligned}
J^{(s)}_j&=\frac{q_j}{\theta}+\frac{\partial \hat{s}_0}{\partial c}J_j^{(m)}+\left(\hat{s}_2(J_{\rho}\rho_{,kk} +J_{c}c_{,kk})+2\rho^2\hat{s}_1v_{k,k}\right)\rho_{,j}\\
&+\left(2\hat{s}_3(J_{\rho}\rho_{,kk}+J_{c} c_{,kk})+\rho^2\hat{s}_2v_{k,k}\right)c_{,j},
\end{aligned}
\end{equation}
where the relation (\ref{solentropy}) has been used.

In the entropy flux (\ref{entropyflux}) we recognize the contribution of the classical term, the ratio between the heat flux and the absolute temperature of the whole mixture, and an entropy extra-flux \cite{Mul} depending on the divergence of the barycentric velocity, together with the first and second order gradients of the mixture mass density  and concentration of one constituent.

The matrix $B^{(2)}_{qr}$, with the above results, identically vanishes, and
the entropy inequality reduces to 
\begin{equation}\label{finalineq}
\begin{aligned}
&q_j\frac{\partial}{\partial x_j}\left(\frac{\partial \hat{s}_0}{\partial\varepsilon}\right)+\left(\tau_6 (v_{k,k})^2+\tau_{12} L_{ij}L_{ij}\right)\frac{\partial \hat{s}_{0}}{\partial\varepsilon} +2\hat{s}_3\frac{(J_{\rho}\rho_{,kk}+J_{c} c_{,kk})^2}{J_{c}}\\
&+J_{\rho}\frac{\partial^2 \hat{s}_0}{\partial\rho\partial c}\rho_{,k}\rho_{,k}+2 J_{\rho}\frac{\partial^2 \hat{s}_0}{\partial c^2}\rho_{,k}c_{,k} +J_{c}\frac{\partial^2 \hat{s}_0}{\partial c^2}c_{,k}c_{,k}\ge 0.
\end{aligned}
\end{equation}
The residual entropy inequality (\ref{finalineq}) turns out to be a homogeneous quadratic polynomial in some gradients entering the state space, whose coefficients depend at most on the field variables; such a relation is fulfilled for all the thermodynamical processes if and only if the following inequalities hold true:
\begin{equation}\label{ineqfinal}
\begin{aligned}
&q \frac{\partial^2 \hat{s}_{0}}{\partial \varepsilon^2}\geq 0,\qquad J_{c}\frac{\partial^2 \hat{s}_{0}}{\partial c^2}\geq 0,\qquad J_{c}\hat{s}_3\geq 0,\qquad \tau_6\frac{\partial \hat{s}_{0}}{\partial \varepsilon}\geq 0,\qquad \tau_{12}\frac{\partial \hat{s}_{0}}{\partial \varepsilon}\geq 0.
\end{aligned}
\end{equation}
Due to conditions \eqref{solentropy}-\eqref{maxentropy} and
\begin{equation}
\begin{aligned}
&\frac{\partial \hat{s}_{0}}{\partial \varepsilon}=s_{01}^{\prime}=\frac{1}{\theta}>0,\qquad\frac{\partial^2 \hat{s}_{0}}{\partial \varepsilon^2}=s_{01}^{\prime\prime}=
-\frac{1}{\theta^2}\theta^{\prime}<0,\qquad\frac{\partial^2 \hat{s}_{0}}{\partial c^2}=\frac{\kappa_1^2}{J_{c}}s_{02}^{\prime\prime},
\end{aligned}
\end{equation}
it follows that the inequalities \eqref{ineqfinal} are satisfied if
\begin{equation}
\label{final}
q\le 0, \qquad J_{c}\le 0,\qquad s_{02}^{\prime\prime}\geq 0,\qquad
\tau_6\ge 0,\qquad\tau_{12}\ge 0; 
\end{equation}
as one expects, the viscosity coefficients, $\tau_6$ and $\tau_{12}$, to be physically meaningful, are non-negative. 

Finally, for the three-dimensional case of a binary non-reacting mixture  that behaves as a viscous Korteweg fluid, described by the mass density of the mixture, the concentration of one constituent, the barycentric velocity and the specific internal energy of the whole mixture, all the thermodynamical restrictions arising from the exploitation of the entropy principle are solved. Furthermore, all the results above detailed allow us to characterize the constitutive functions involved in the model, and complete the proof.
\end{proof}

As a final remark, let us check how the model we considered may be useful to investigate phase transitions \cite{Fabrizio-Giorgi-Morro-2006,Lowengrub-Truskinovsky,Heida-Malek-Rajagopal,morro2006,morro2007}; as a consequence, we derive an evolution equation for a phase-field (the concentration $c$, in our setting) encompassing the classical Cahn-Hilliard equation. To this end, by restricting ourselves to a situation not too far from equilibrium, we consider a semilinear approximation of Eq.~(\ref{modelequations})$_2$, \emph{i.e.}, we
neglect the product of any couple of spatial or time derivatives \cite{COP-2016}.

 Let us recall  the classical Cahn-Hilliard equation \cite{Cahn-Hilliard},
\begin{equation}\label{9}
\dot c+k\alpha c_{,kkqq}=k[f^{\prime}(c)]_{,qq},
\end{equation}
where $k$ and $\alpha$ are two constants, $f$ is the double-well free energy due to the presence of a diffuse interface, and the superposed dot denotes the material time derivative, \emph{i.e.}, $\dot c= c_{,t}+ c_{,j}v_{j}$.

In a semilinear approximation, the right-hand side of Eq.~(\ref{9}) becomes
\begin{equation}\label{10}
k[f^{\prime}(c)]_{,qq}=kf^{\prime\prime\prime}(c)c_{,q}c_{,q}+kf^{\prime\prime}(c)c_{,qq}\approx kf^{\prime\prime}(c)c_{,qq},
\end{equation}
whence we get
\begin{equation}\label{11}
k c_{,qq}\approx\frac{k[f^{\prime}(c)]_{,qq}}{f^{\prime\prime}(c)}.
\end{equation}
On the other hand, under the same approximation, we may assume
\begin{equation}\label{12}
(\rho\dot c)_{,qq}\approx\rho\dot{\overline{ c_{,qq}}}.
\end{equation}
In agreement with the approaches developed in \cite{Lowengrub-Truskinovsky,Heida-Malek-Rajagopal}, let us suppose the specific entropy to be dependent only on the square of the concentration gradient; then, choosing $\kappa_1=\kappa_2=0$, Eq.~(\ref{modelequations})$_2$ reads
\begin{equation}\label{13}
\rho\dot{c}+\kappa_3c_{,jj}=0,
\end{equation}
and the diffusional mass flux reduces to
\begin{equation}
J_j^{(m)}=\kappa_3 c_{,j}\qquad \hbox{with}\quad \kappa_3\leq 0.
\end{equation}
Now, taking the laplacian of Eq.~(\ref{13}), in the semilinear approximation, provides
\begin{equation}\label{15}
\rho\dot{\overline{ c_{,qq}}}+\kappa_3c_{,jjqq}=0.
\end{equation}

Finally, combining equations (\ref{13}) and (\ref{15}), a general semilinear partial
differential equation ruling the evolution of a phase-field which characterizes the phase transitions in binary mixtures of third grade Korteweg fluids is recovered, say
\begin{equation}\label{18}
\rho\dot{c}+\kappa_3c_{,jjqq}=-\kappa_3c_{,jj} - \rho\dot{\overline{ c_{,qq}}}.
\end{equation}
The evolution equation \eqref{18} for the phase-field is more general than the classical Cahn-Hilliard equation. Nevertheless,
under suitable hypotheses, it may be reduced to the classical Cahn-Hilliard equation.
We observe that this procedure can be exploited by considering first order spatial or time derivatives of the phase-field, under the additional hypothesis that second order quantities are negligible. From the physical point of view, this corresponds to consider slow transitions (small $c_{,t}$) with moderate inhomogeneity (small $c_{,j}$) \cite{COP-2016}.

\section{Conclusions}
\label{sec:final}
In this paper, we considered a non-reacting binary mixture of viscous third grade fluids with one temperature, and analyzed the compatibility of the constitutive functions with
the second law of thermodynamics. Because of the non-locality of the constitutive equations, we exploited the entropy principle by means of the extended Liu procedure, which consists in taking into account the field equations and their gradient extensions. 
Firstly, we introduced the basic fields to describe the mixture as a whole, \emph{i.e.}, the mass density of the mixture, the concentration of one constituent, the barycentric velocity, and the specific internal energy of the mixture and, then, the related governing equations. 
The studied model neglects the internal exchange of mass due to chemical reactions, the action
of external body forces (say, the gravity), and heat sources as well as
momentum and energy exchanges between the components. 

We assumed a state space containing the first order gradients of the basic fields, and the second order gradients of mixture mass density and concentration.
Therefore, by applying the procedure, we were able to characterize the non-local constitutive equations guaranteeing the compatibility with second law of thermodynamics. 
All the long, though straightforward, computations have been managed by using some symbolic routines written in the Computer Algebra System Reduce \cite{Reduce}. 

The present model  exhibits a smaller level of detail with respect to the case with two temperatures analyzed  in \cite{CGOP-2020,GORbis-2021}; nevertheless, the diffusive effects are described by a diffusional mass flux recovered from the thermodynamic restrictions.

As a consequence of the algorithmic procedure, after expanding around the equilibrium the specific entropy, retaining only first order terms in the gradients of the mixture mass density and concentration of one constituent, the expression of the entropy flux naturally arises. The latter contains the classical part and an extra-flux depending on the first and second order gradients of mixture mass density and concentration, as well as on the divergence of barycentric velocity.

It is worth of being remarked that we did not assume \emph{a priori} neither a modified
energy equation including additional contributions (\emph{i.e.}, like interstitial working \cite{DunSer}),
nor postulate a generalized expression of the entropy flux as often done in non-local ﬂuid mechanical  \cite{morro2007}, or phase-ﬁeld
models \cite{Heida-Malek-Rajagopal,morro2006,Fabrizio-Giorgi-Morro-2006}.

Remarkably, we were able to explicitly solve the thermodynamical constraints by obtaining a Cauchy stress tensor that is sufficiently general to 
include that proposed by Korteweg in \cite{Kor}, and prove that the mixture's behavior is typical of a viscous Korteweg fluid \cite{GP-2021}. Finally, by means of a semilinear approximation, we derived a generalized evolution equation for a phase-field characterizing phase transitions in binary mixtures of third grade Korteweg fluids.

The theoretical results above derived contain some degrees of freedom and may serve as a basis for experimental and/or numerical investigations. 
Further applications of the theory could be the investigation of diffusive binary mixtures of Korteweg-type fluids with one temperature, where the diffusion velocity is included in the state space; as a consequence, the diffusional mass flux will not represent a constitutive function, and the thermodynamical constraints involve a contribution due to the diffusion velocity.

\section*{Acknoledgments}
Work supported by G.N.F.M. of the Istituto Nazionale di Alta 
Matematica ``F. Severi''. M.G. acknowledges the support through the ``Progetto Giovani GNFM 2020''. The authors thank the unknown Referees for the
helpful comments leading to clarify some aspects and improve the quality of the paper.

\end{document}